\documentstyle[aps,preprint,pre]{revtex}
\begin{document}
\date{\today}
\draft
\title{Phase transitions in interacting domain-wall model}
\author{Jae Dong Noh and Doochul Kim}
\address{Department of Physics and Center for Theoretical Physics,
Seoul National University,\\ Seoul 151--742, Korea}
\maketitle
\begin{abstract}
We investigate the interacting domain-wall model derived from the 
triangular-lattice antiferromagnetic
Ising model with two next-nearest-neighbor interactions.
The system has commensurate phases with a domain-wall density $q=2/3$ 
as well as that of $q=0$ when the interaction is repulsive. The
$q=2/3$ commensurate phase is separated from the incommensurate phase 
through the Kosterlitz--Thouless~(KT) transition. 
The critical interaction strength and the nature of the KT 
phase transition are studied by the Monte Carlo simulations 
and numerical transfer-matrix calculations. For  strongly attractive 
interaction, the system undergoes a first-order phase transition from 
the $q=0$ commensurate phase to the incommensurate phase with $q\neq
0$. The incommensurate phase is a critical phase which is in the 
Gaussian model universality class. The effective Gaussian coupling 
constant is calculated as a function of interaction parameters from 
the finite-size scaling of the transfer matrix spectra .
\end{abstract}
\pagebreak

\section{Introduction}\label{sec1}
The triangular-lattice antiferromagnetic Ising model~(TAFIM) 
displays rich critical phenomena. The TAFIM with 
nearest-neighbor coupling $K~(<0)$ is described by the Hamiltonian
\begin{equation}\label{H.tafim}
-\frac{H}{k_B T} = K \ \sum_{\langle ij\rangle}\  s_i\  s_j
\end{equation}
where $\langle ij\rangle$ denotes pairs of nearest-neighbor
sites of triangular lattice and $s_i=\pm 1$.
The ground states of the model are infinitely degenerate due to 
frustration on each elementary triangle and correspond to a critical 
state with algebraic decay of correlations~\cite{BloH82}. Each 
ground-state configuration can be mapped to a state of the `triangular
solid-on-solid~(TISOS)' model~\cite{NieHB84} which describes 
equilibrium shape of a simple-cubic crystal near its $[1,1,1]$ corner 
or growing of the simple-cubic crystal along the $[1,1,1]$ 
direction~\cite{ForT90}, and also of the  domain-wall 
model~\cite{BloH82,NohK94}. The latter describes the 
commensurate-incommensurate (C-IC) phase transitions~\cite{Bak82}.
When a two-dimensional system with anisotropic interactions
has degenerate ground states, excitations from a ground state
may take the form of domain walls of striped shapes. A
typical example is the axial next-nearest neighbor Ising (ANNNI) 
model~\cite{VilB81}. Though
domain-wall type excitations are energetically unfavorable, they also
carry a finite amount of entropy. So the system is in an ordered 
phase without domain-wall excitation (C phase) at low temperatures 
while it is in a modulated phase with finite density of domain wall 
for sufficiently high temperatures.

Ground-state properties of the TAFIM are studied by taking the limit
$K\rightarrow -\infty$ in Eq.~(\ref{H.tafim}) and by assigning
statistical weights to each ground-state configuration. 
The new effective Hamiltonian of interest here can then be written 
as
\begin{equation}\label{eff_hamil}
{\cal H} = -\sum_{\langle ij\rangle} \delta_a s_is_j - 
\sum_{\langle\!\langle ij\rangle\!\rangle} \epsilon_a s_i s_j - 
h \sum_{i} s_i \ \ \ .
\end{equation}
Here, $\langle ij\rangle$ ($\langle\!\langle ij\rangle\!\rangle$) 
denotes pairs of nearest-neighbor (next-nearest-neighbor) sites, 
$\delta_a$~($\epsilon_a$), $a=1,2,3$, are direction-dependent
nearest-neighbor~(next-nearest-neighbor) couplings and $h$ is a 
magnetic field. Each allowed spin configuration $\{s_i\}$ is 
understood to be one of the ground-state configurations implicitly.
The statistical weight for each configuration is proportional to
$e^{-{\cal H}[\{s_i\}]}$.

If there are only nearest-neighbor couplings, the system is 
equivalent to the non-interacting domain-wall model and can be solved 
exactly~\cite{BloH82,NohK94}. The system exhibits 
Pokrovsky-Talapov~(PT) transition~\cite{PokT79} which separates an 
ordered commensurate~(C) phase and a disordered incommensurate~(IC)
phase. The IC phase is a critical phase which is in the Gaussian
model universality class. The free-energy functional of the Gaussian
model is given by
\begin{equation}\label{ftnal_G}
{\cal F} = \int d^2{\bf r} \left[\,\frac{K_1}{2} \left( 
\frac{\partial\phi}{\partial x_1} \right)^2 +\frac{K_2}{2} \left( 
\frac{\partial\phi}{ \partial x_2}\right)^2 \right]
\end{equation}
where $K_1 \mbox{ and }K_2$ are the anisotropic stiffness constants. 
The Gaussian model has scaling dimensions
\begin{equation}\label{sc_dim}
X_{p,q} = \frac{p^2}{2g}+\frac{g}{2}q^2
\end{equation}
where $p$ and $q$ are integers and $g=2\pi\sqrt{K_1K_2}$ is called 
the Gaussian coupling constant~\cite{VilB81}. The Gaussian coupling 
constant is equal to $1/2$ throughout the IC phase in this 
limit~\cite{NohK94}.

Effect of the magnetic field for the system with $\delta_a= 
\epsilon_a=0$ has been studied by numerical transfer-matrix
calculations~\cite{BloNWH91} and Monte Carlo 
simulations~\cite{BloN93}. The system undergoes 
Kosterlitz--Thouless~(KT) phase transition~\cite{xykt} from the 
disordered critical phase to the ordered phase as $h$ increases.

When there are two next-nearest-neighbor couplings~($\epsilon_1\neq
0,\epsilon_2\neq 0,\epsilon_3=0$) as well as nearest-neighbor 
couplings, ${\cal H}$ in Eq.~(\ref{eff_hamil}) can be interpreted as 
a Hamiltonian for an interacting domain-wall system~\cite{NohK94}. 
The nearest-neighbor and next-nearest-neighbor 
couplings play the role of the fugacity of domain walls and the 
interaction between domain walls, respectively. There are two types 
of domain walls and the two next-nearest-neighbor couplings 
$\epsilon_1$ and $\epsilon_2$ control wall-wall interactions between 
respective types.  In the previous paper~\cite{NohK94}, we 
considered the partially interacting domain-wall~(PIDW) model where 
domain walls of only one type interact with each 
other~($\epsilon_1=0,\ \epsilon_2\neq 0$ case).
The PIDW model is equivalent to the general five-vertex model which 
in turn is exactly solvable through the Bethe Ansatz method. 
We obtained exact phase diagram and quantitative understanding
of the effect of the interaction in the critical properties of the
PIDW model. When the repulsive interaction is strong, the system is 
in a novel C phase with domain-wall density $q=1/2$
which is absent in the non-interacting system. 
In fact, this phase is equivalent to the C phase appearing in the
ANNNI model where the interactions between domain walls are assumed 
to be infinitely repulsive~\cite{VilB81}. 
The IC phase is also the critical phase in
the Gaussian model universality class but the Gaussian coupling 
constant $g$ varies continuously throughout the IC phase.

In this paper, we study the interacting domain-wall~(IDW) 
model where domain walls of both types interact with
same strength~($\epsilon_1=\epsilon_2\neq0$ case).
We are concerned with the nature of the ordering and the phase 
transition. We find that the system has an ordered C phase with 
$q=2/3$ when the repulsive interaction is sufficiently strong and 
suggest that the phase transition between the
IC phase and the $q=2/3$ C phase is the KT transition. We argue 
that the transition is described in the continuum limit by the
Gaussian model Eq.~(\ref{ftnal_G}) with a symmetry-breaking field $V_3
\cos\left(3\phi({\bf r})\right)$. 
The repulsive interaction acts as the symmetry-breaking field.
When $V_3$ is sufficiently small, the system
is in the critically disordered state with scaling dimensions given 
in Eq.~(\ref{sc_dim}). The Gaussian coupling constant is not constant 
any more in the critical region but varies continuously since it is 
renormalized by the $V_3$ field. The symmetry-breaking field $V_3$ 
has a scaling dimension $X_{3,0}$.
As $V_3$ increases, it becomes relevant in the renormalization group 
sense. This happens when $X_{3,0}=2$ (or $g=9/4$).
The phase transition involved in this order-disorder transition
is the KT transition. So, we predict that the KT transition occur as 
the repulsive-interaction strength varies. 

On the other side, where the attractive interaction is strong, the 
domain walls are bounded with each other. We find that, as the 
chemical potential of domain walls increases, there is a first-order 
phase transitions from the $q=0$ C phase to the IC phase accompanied
by a finite jump in $q$.

This paper is organized as follows. In Sec.~\ref{sec2} we introduce 
the IDW model and present the relation between it and the TISOS 
model. Possible ordered phases in the IDW model are discussed. 
Unlike the PIDW model, the IDW model is not exactly solvable.
So, we investigate the IDW model by numerical diagonalization of 
the transfer matrix and Monte Carlo simulations. In Sec.~\ref{sec3} 
we analyze the eigenvalue spectra of the transfer matrix using the
finite-size scaling theory. The transfer matrix applies to a system 
with a cylinder geometry: infinitely long in one direction, and 
periodic with a finite size $N$ in the other direction. 
The calculations are performed for the strip width $N$ up to $N=18$, 
corresponding to a transfer matrix of size $2^{18}\times 2^{18}$.
Using the finite-size scaling of the eigenvalue spectra, we 
calculate the Gaussian coupling constant $g$ from which the KT 
transition point for the repulsive interaction is located.
The first-order phase transition line for the attractive interaction 
is obtained from the eigenvalue spectra and an analytic 
approximation.  In Sec.~\ref{sec4} results of Monte Carlo simulations 
are presented. The simulations are done for a lattice of size up to 
$120\times 120$ while the domain-wall density is fixed to $2/3$. 
As well as the specific heat, we also calculate the fluctuations of
the position and the density of domain walls since they serve as 
order parameters for the KT transition. Monte Carlo results are 
consistent with those of transfer-matrix calculations.
In Sec.~\ref{sec5} we summarize our results and discuss the origin of 
the difference between the PIDW model and the IDW model.

\section{The model and its ground states}\label{sec2}
We consider two-dimensional domain-wall model on a 
finite $N\times M$ triangular lattice with one side of elementary
triangle lying along the horizontal direction. The configurations of 
domain walls are mapped from the ground-state configurations of the 
TAFIM~\cite{NohK94}.
The domain walls propagate vertically along the two directions of 
elementary triangle. We call the domain wall of type 1 (2) if it is 
left (right) moving. A single chain of domain wall has an entropy 
$\log{2}$ per unit length since there are two ways for domain walls 
to move. The configurations of domain walls are
subject to the restriction that they never be created or annihilated
during propagation. Thus, we do not allow the `dislocations' 
which form topological excitations. However, effect of such 
excitations will be discussed briefly in Sec.~\ref{sec5}.

Domain walls have the chemical potential $\mu$ and interaction 
energy $\varepsilon$ per unit length. Relationships between
$(\mu,\varepsilon)$ and $(\delta_a,\epsilon_a)$ of
Eq.~(\ref{eff_hamil}) are derived in 
Ref.~\cite{NohK94}. The interaction energy is assigned to
each segment of adjacent-parallel domain walls.
The grand-canonical partition function of the IDW model is written 
as
\begin{equation}\label{gcpf}
{\cal Z}_G(x,y) =  \sum_{Q=0}^N x^{MQ} \sum_{\{p_1,\ldots,p_Q\}} 
y^{{\bf L}(p_1,\ldots,p_Q)}
\end{equation}
where $Q$ is the number of domain walls per row, $x=e^{\mu}$ is the
fugacity of a domain wall segment of unit length and 
$y=e^{-\varepsilon}$ is the interaction parameter. Here, we use the
convention that the `temperature' of the domain-wall system is absorbed
into $\mu$ and $\varepsilon$. Note that this temperature is different
from that of the TAFIM.
The function ${\bf L}(p_1,\ldots,p_Q)$ counts total length of 
adjacent-parallel domain walls where $p_i(j)$ denotes the horizontal 
position of the $i$-th domain wall at vertical position $j$. 
Since configurations with different number of domain walls do not 
couple with each other at all, the partition sum can be evaluated 
independently for each $Q$;
$$
{\cal Z}_G = \sum_{Q=0}^N x^{MQ}\  {\cal Z}_C(y,q)
$$
where ${\cal Z}_C$ is the canonical partition function,
representing the second sum in Eq.~(\ref{gcpf}), and $q=Q/N$ is the
domain-wall density. The grand-canonical and canonical ensembles are 
connected through the relation
\begin{equation}\label{q_mu}
\mu = -\frac{1}{MN} \frac{\partial \log{\cal Z}_C}{\partial q}\ \ \ .
\end{equation}

Before proceeding the analysis of the IDW model, we digress to 
explain the equivalence of the domain-wall model and the TISOS model. 
The TISOS model is a solid-on-solid model on a triangular lattice on 
which integer-valued heights $h_i$ at site $i\equiv (x_i,y_i)$ are 
assigned with the restriction that the height differences between any 
nearest-neighbor sites are $\pm 1$ or $\pm 2$. 
See Ref.~\cite{BloH82} for details.
Let $n_i$ be the number of domain walls on the left-hand side of the 
point $(x_i,y_i)$. Then, the mapping
\begin{equation}\label{ht_n}
h_i = 3\,n_i - 2\,x_i
\end{equation}
relates a domain-wall configuration to a height configuration.
Note that the height variables satisfy the periodic boundary 
condition $h_{(x_i+N,y_i)}=h_{(x_i,y_i)}$ only if the domain-wall 
density $q=2/3$.
In this case the TISOS model so constructed, describes the 
equilibrium shape of a cubic crystal viewed from the $[1,1,1]$ 
direction. The case of $q\neq 2/3$ corresponds to that viewed from 
other directions. 

`Ground states' of the model for given $x$ and $y$ are the
configurations which maximize the Boltzmann weight in 
Eq.~(\ref{gcpf}). For the attractive interaction~$(\varepsilon<0)$, 
a C phase with $q=0$ $(1)$ is the ground state when 
$\mu<\varepsilon$ $(\mu>\varepsilon)$. There will be a first-order 
phase transition along the line $\mu=\varepsilon$ at zero 
temperature, i.e., $|\varepsilon|\rightarrow\infty$. 
On the other hand, for the repulsive 
interaction $(\varepsilon>0)$, $q=0(1)$ C phase is the ground state 
when $\mu<0(\mu>3\varepsilon)$. When $0<\mu<3\varepsilon$, the 
$q=2/3$ C configurations become the ground states. These new ground 
states correspond to the most densely-packed domain-wall states 
without interacting pairs and are three-fold degenerate. 
They are shown in Fig.~1.

To understand finite-temperature behaviors, we should consider 
excitations from these ground states. Excited states from the 
$q=0$ state are formed by domain-wall formations. 
If the domain walls are far apart from each other (this is the case 
when the attractive interaction is not so strong), the energy cost 
is $-\mu$ and the wandering entropy gain is $\log 2$ per unit length 
of domain walls. Thus there will be a phase transition at 
$\mu=-\log 2$.

An excited state of the $q=1$ state is formed  by removing one 
segment of domain walls from each row. We will call it the 
domain-wall hole excitation. Unlike a domain wall, the domain-wall 
hole need not form a continuous path. Thus a single domain-wall 
hole has a wandering entropy $\log N$ and takes energy cost $\mu$ 
per unit length. So the $q=1$ state is always unstable under 
domain-wall hole excitations at any value of $\mu$.

The $q=2/3$ C state has complicated excitations. The simplest kind of 
excitations is generated by locally reversing two units of a domain 
wall with respect to a vertical axis. Its excitation energy is 
$2\varepsilon$.  This excitation is a local excitation so it cannot 
destroy the long-range order at low temperatures. Another important 
excitation is formed by domain boundaries between the three 
ground-state configurations. This is illustrated in Fig.~2 where 
three ground-state configurations of Fig.~1 coexist forming domain 
boundaries. 

Nature of such excitations becomes more transparent when we 
introduce a coarse-grained field variable as follows. First we assign 
a height variable $h_i$ to each lattice site according to 
Eq.~(\ref{ht_n}). The height variables corresponding to the domain 
wall configuration of Fig.~2 are also shown in Fig.~2 by small 
numbers. We now group lattice sites into a triangular array of 
elementary triangles shown as dotted triangles in Fig.~2. To each 
elementary dotted triangle $\alpha$, we then assign a
coarse-grained height variable
\begin{equation}\label{cgh}
\tilde{h}_\alpha \equiv (h_{\alpha_1}+h_{\alpha_2}+h_{\alpha_3}) / 3
\end{equation}
where $\alpha_i$'s denote the three sites of the triangle $\alpha$. 
The values of $\tilde{h}_\alpha$ for the configuration of Fig.~2 
are shown by big numbers in the middle of each dotted triangle.
$\tilde{h}_\alpha$ take integer values and $\tilde{h}_\alpha=
\tilde{h}_{\alpha '}\ (\mbox{mod\ } 3)$ if local configurations at 
$\alpha$ and $\alpha '$ are in the same ground-state configuration. 
At each boundary, $\Delta \tilde{h}_\alpha =\pm 1\mbox{\ or\ } 
\pm 2$.  In the continuum limit, the critical behaviors of the 
system are described by the Gaussian model with free-energy 
functional ${\cal F}$ in Eq.~(\ref{ftnal_G}) if there is no 
interaction between domain walls.  
The repulsive interaction favors the coarse-grained height 
$\tilde{h}({\bf r})$ to be an integer where $\tilde{h}({\bf r})$
denotes $\tilde{h}_\alpha$ in the continuum limit.
So we postulate that the order-disorder phase transition in the IDW 
model with $q=2/3$ is described by the Gaussian model with a 
symmetry-breaking field whose free-energy functional is given by
\begin{equation}\label{feftnal}
{\cal F} = \int d^2\!{\bf r}\  \left[ \frac{1}{2}\,K_1 
\left(\frac{\partial \phi}{\partial x_1}\right)^2 + 
\frac{1}{2}\,K_2 \left(\frac{\partial \phi}{ \partial x_2}\right)^2 + 
V_3 \cos\left( 3\phi({\bf r})\right) \right]
\end{equation}
where $\phi({\bf r})=\frac{2\pi}{3}\tilde{h}({\bf r})$ and the 
parameters $K_1,K_2$ and $V_3$ are functions of the interaction 
strength $y$.  It is known that this system has two phases: 
critically disordered phase (high temperature phase) and the 
ordered phase (low temperature phase)~\cite{xykt}. 
In the critical region the model is renormalized to
\begin{equation}\label{feftnalr}
{\cal F} = \int d^2\!{\bf r} \left[\ \frac{1}{2}\,\widetilde{K}_1 
\left(\frac{\partial \phi}{\partial x_1}\right)^2 + 
\frac{1}{2}\,\widetilde{K}_2 
\left(\frac{\partial \phi}{ \partial x_2}\right)^2 \right]
\end{equation}
where $\widetilde{K}_1$ and $\widetilde{K}_2$ are renormalized 
stiffness constants which are functions of $K_1,K_2$ and $V_3$. The
scaling dimension of the symmetry-breaking field is $X_{3,0}=9/(2g)$
with the effective Gaussian coupling constant $g=2\pi\sqrt{
\widetilde{K}_1 \widetilde{K}_2}$. In the critical region the 
symmetry-breaking field is irrelevant (i.e., $X_{3,0}>2$). The phase 
transition into the ordered phase occurs at the values of the model
parameters which satisfy the relation $X_{3,0}=2$ and belongs to the 
universality class of the KT transition. The transition point can be 
found from the condition $g=9/4$.

In this section, we presented ground states of the IDW model and 
the conjecture that the C-IC transition 
of the IDW model with the repulsive interaction at $q=2/3$ is 
described by the free-energy functional given in Eq.~(\ref{feftnal}). 
In the next two sections, we will confirm our conjecture by transfer
matrix calculations and Monte Carlo simulations.

\section{Finite-size scaling analysis of transfer-matrix 
spectra}\label{sec3}
We study the IDW model on a finite lattice with width $N$ and height 
$M$ using the transfer matrix. The grand-canonical partition function
${\cal Z}_G$ in Eq.~(\ref{gcpf}) is written as
$$
{\cal Z}_G = \sum_{Q=0}^{N} \ x^{MQ}\ {\rm Tr}\ {\bf T}_Q^M
$$
where ${\bf T}_Q$ is a transfer matrix which acts on the ${}_NC_Q$
dimensional space spanned by statekets $|n_1,\ldots,n_Q\rangle$
specifying positions of $Q$ domain walls. 
The matrix element $\langle n_1',\ldots,n_Q'\mid {\bf
T}_Q\mid n_1,\ldots,n_Q\rangle$ represents the Boltzmann weight for a
domain-wall configuration between two rows of a lattice as shown in
Fig.~3.

We denote the $p$-th largest eigenvalue of ${\bf T}_Q$ by 
$e^{-E_Q^p}$ and $E_Q^p$ will be called the energy of the {\it p}-th 
excited level.  In the limit $M\rightarrow \infty$, the canonical 
partition function is given by the largest eigenvalue, i.e., 
${\cal Z}_C = e^{-M E_Q^0}$. In the IC phase, $q=Q/N$ varies
continuously as a function of $\mu$ and $\varepsilon$. When $q$ is
given, the chemical potential should be chosen so as to satisfy the
finite-size version of Eq.~(\ref{q_mu})
$$
\mu = \frac{E_{Q+1}^0-E_{Q-1}^0}{2N}\ \ \ .
$$

The eigenvalue spectrum for a conformally invariant system
follows the universal scaling form
\begin{equation}\label{fss}
{\rm Re}\ E_\alpha(N)  =  N f_\infty+
\frac{2\pi\zeta}{N} \left(X_\alpha-\frac{c}{12}\right) + 
o\left(\frac{1}{N}\right)
\end{equation}
where $c$ is the central charge, 
$X_\alpha$ is the scaling dimension of the operator associated with 
eigenstate $\alpha$, $\zeta$ is an anisotropy factor, and $f_\infty$
is the bulk free energy per site~\cite{KimP87}.  
The IC phase of the domain-wall system is the critical phase in the
universality class of the Gaussian model whose central charge is $1$
and scaling dimensions are given in Eq.~(\ref{sc_dim}).
In the previous paper~\cite{NohK94}, we showed that
the $p$-th excited level of ${\bf T}_Q$ leads to the scaling 
dimension $X_{p,0}$ and excitations of the largest eigenvalues of 
${\bf T}_{Q\pm q}$ relative to that of  ${\bf T}_Q$ lead to 
$X_{0,q}$.
Using this knowledge, we can estimate the Gaussian coupling constant 
by an estimator $g(N)$ defined by
$$
g(N) = \sqrt{\frac{ {\rm Re}\ (E_{Q+1}^0+E_{Q-1}^0-2\,E_Q^0)}{ 2\,
{\rm Re}\ (E_Q^1-E_Q^0)} }
$$
which should approach $g$ as $N\rightarrow\infty$. A sequence of 
$g(N)$ is obtained by numerical diagonalization of transfer matrix 
for a system of width $N$ up to 18. We present the estimates in 
Fig.~4(a) and Fig.~4(b) for the repulsive~$(y<1)$ and 
attractive~$(y>1)$ case, respectively for several values of $y$.
The curve for $y=0.2$ in Fig~.4(a) is noticeable. $g(N)$ apparently 
does not converge to a finite value as $N$ increases when $q=2/3$. 
It means that when $y=0.2$ the eigenvalue spectrum does not follow 
the finite-size scaling form of Eq.~(\ref{fss}) and that the system 
is no longer in a critical phase.  
We gave an argument in the previous section that the system undergoes 
a phase transition into the ordered phase as the repulsive 
interaction becomes strong and that the Gaussian coupling constant 
should be  $9/4$ at the transition point. Thus we estimate the 
critical interaction strength $y_c$ by solving the equation 
$g(N)=9/4$ numerically. For this purpose, we find that the alternative
estimators defined by 
$$
\tilde{g}(N) = -\frac{6\left( E_Q^0(N+3)-E_Q^0(N) \right)}{ \left(
E_Q^1(N+3)-E_Q^0(N+3)\right)-\left(E_Q^1(N)-E_Q^0(N)\right)}
$$
converge better. The solutions of $\tilde{g}(N)=9/4$, denoted by 
$y_1(N)$, are given in Table~I.  
The strip-width dependence of $y_1(N)$ follows from corrections
to scaling of the eigenvalues. At the KT transition point there
are logarithmic corrections due to the presence of a marginally
irrelevant field~\cite{BloN93}. Thus the sequence of estimator 
$y_1(N)$ is extrapolated by fitting it to the form 
$y_1(N) = y_2+ a/(b+\log N)$. Since the sequence $y_2$ thus obtained
also show slow convergence, we fit it again to the same form and 
obtain the result $y_c=0.252\pm(0.003)$. See Table I.

When $y<y_c$, there appears the $q=2/3$ C phase. As in the $q=1/2$ C
phase of the PIDW model, $q$ remains locked to the value $2/3$ in the 
range $\mu_-\leq \mu \leq \mu_+$. Estimates of the chemical potential 
at the lower and upper boundaries of the C phase are given by
$$
\mu_\pm(N) = \mp \left( E_{Q_0\pm 1}^0(N) - E_{Q_0}^0(N) \right)
$$
where $Q_0=2N/3$. Since they are evaluated from finite-size systems, 
we should know the finite-size scaling form of $\mu_\pm(N)$. 
Due to the marginal corrections at the phase boundary, the largest 
eigenvalues for $Q=Q_0\pm m$ is expected to have scaling forms
$$
E_{Q_0 \pm m}^0 = E_{Q_0}^0-\mu_\pm m +\frac{\pi\zeta g}{N} m^2 +
O\left(\frac{1}{N\log N}\right) \ \ \ .
$$
It then follows that $\mu_\pm(N)$ behaves as 
$$
\mu_\pm(N) = \mu_\pm \mp \frac{\pi\zeta g}{N} + 
O\left(\frac{1}{N\log N}\right) \ \ \ .
$$
Accordingly, we estimate $\mu_\pm$ by fitting $\mu_\pm(N)$ to the 
equation $\mu_\pm(N)=\mu_\pm+a/N+b/N\log N$. The result is shown in 
Fig~5.

We have thus determined the phase boundary of the $q=2/3$ phase using 
the finite-size scaling of the eigenvalue spectra of the transfer
matrix. Next, we consider the system near the $q=0$ C phase. As the
chemical potential $\mu$ increases, domain walls begin to form. They
form a free or bound state depending on the
interaction strength $y$. To see this, we calculate in the Appendix 
the ground-state energy and un-normalized eigenket of ${\bf T}_Q$ 
for $Q=2$ from the Bethe Ansatz method. The results for the maximum 
eigenvalue and the eigenket are 
\begin{equation}\label{vv2}
\begin{array}{lllll}
E_2^0 & = & -2\log 2  ,& |\lambda\rangle = \sum_{m<n} |m,n \rangle &
\ \ \ \mbox{if\ \ \ } y\leq \frac{3}{2} \\
E_2^0 & = & -\log\left( 2y+\frac{1}{2(y-1)}\right) ,& 
|\lambda\rangle =
\sum_{m<n} \left(2(y-1)\right)^{-(n-m)} |m,n\rangle &\ \ \  
\mbox{if\ \ \ } y\geq\frac{3}{2} \ \ \ ,
\end{array}
\end{equation}
where the thermodynamic limit is taken.
The eigenket has a qualitatively different properties on both side of
$y=3/2$ line. 

As can be seen from Eq.~(\ref{vv2}), domain walls form a free state 
when $y\leq 3/2$. So, the C-IC transition is the PT transition as in 
the non-interacting and the PIDW models. The domain wall density has 
a square-root dependence on the chemical potential
$$
q \sim (\mu-\mu_c)^{1/2}
$$
with the critical chemical potential $\mu_c = -\log 2$.

When $y\geq 3/2$, Eq.~(\ref{vv2}) shows that domain walls form a 
bounded state. The two domain walls are bounded with the mean 
distance
$$
d_0 = \langle m-n\rangle = \frac{4(y-1)^2}{4(y-1)^2-1}.
$$
The ground-state energy of the bounded state consists of two parts 
$E_2^0=E_{\rm c.m.}+E_{\rm int.}$ where $E_{\rm c.m.}=-\log 2$ comes 
from free motions of the center of mass of the two domain walls and 
$E_{\rm int.}=\log\left(y+1/4(y-1) \right)$ from the effective 
interaction. To generalize Eq.~(\ref{vv2}) for $Q\geq 3$ where exact 
results are not available, we construct a simple approximation for 
$E_Q^0$ as a sum of 
$E_{\rm c.m.}$ and $E_{\rm int.}$. Thus we  write the ground-state 
energy of the ${\rm T}_Q$ for $Q\neq 2$ as
\begin{eqnarray}\label{e_gbdw}
E_Q^0 &=& E_{\rm c.m.} + (Q-1) E_{\rm int.} \\ \nonumber
&=& -\log 2 - (Q-1) \log\left(y+\frac{1}{4(y-1)}\right) \ \ \ .
\end{eqnarray}
This approximation will be called the two-domain-wall approximation.
It should be valid as long as $q<\!<1/d_0$. 
When $q{\hbox{\kern1.5mm\raise0.8mm\hbox{$>$}\kern-3.2mm
       \raise-1.0mm\hbox{$\sim$}\kern1.5mm}} 1/d_0$, domain 
walls feel additional statistical repulsive interactions. 
This statistical repulsions then drive the system into the IC phase.
Within the two-domain-wall approximation, the domain-wall system
undergoes a first order C-IC transition from the $q=0$ C phase to the
IC phase. Since Eq.~(\ref{e_gbdw}) is linear in $Q$, the transition
occurs at the chemical potential 
\begin{equation}\label{apx.mu_0}
\mu_0 = -\log\left( y+\frac{1}{4(y-1)}\right)
\end{equation}
while $q_0$, the discontinuity of $q$ at the transition, can be taken
to be $1/d_0$ approximately.

$\mu_0$ and $q_0$ are also determined numerically. 
When $y\geq 3/2$, $E_Q^0$ for finite $N$ obtained from 
numerical diagonalization of ${\bf T}_Q$ becomes concave downward in 
some region $0\leq Q\leq Q_0$ indicating a first order phase 
transition. We took the convex envelop of $E_Q^0$ and estimated 
$\mu_0$ and $q_0$ from  the slope of the straight line enveloping the 
concave region and from the point where the straight line and the 
curve of $E_Q^0$ are tangential, respectively. The results are 
shown in Figs.~6 for five values of $N$. These are compared with
Eq.~(\ref{apx.mu_0}) and the approximation $q_0\sim 1/d_0$ in Figs.~6.

Resulting phase diagram of the IDW model as obtained from the 
transfer-matrix calculations is summarized in Fig.~7. 
This is the main result of this work.

\section{Results of Monte Carlo simulations}\label{sec4}
In the previous section, the assumption that the repulsive 
interaction induces the KT transition at the  $q=2/3$ C phase boundary
is used to 
estimate the critical interaction strength $y_c$ from the numerical 
diagonalization of the transfer matrix.  In this section, we will 
confirm the nature of the phase transition more directly by Monte 
Carlo simulations at $q=2/3$ on a finite $L\times L$ lattice.

At the KT transition point, there is no diverging peak in the
specific heat. So, we should study another quantities such as the 
renormalized stiffness constants $\widetilde{K}_1$ and 
$\widetilde{K}_2$ in Eq.~(\ref{feftnalr}). They can be evaluated 
from the linear response
$$
\left< \frac{\partial\phi}{\partial x_k} \right> =
\ \frac{v_k}{\widetilde{K}_k}
$$
produced by an additional free energy term
$$
\delta {\cal F}_k = -v_k \int d^2\!{\bf r}
\ \frac{\partial\phi}{\partial x_k}
$$
where $k=1\ \mbox{or}\ 2$. From the fluctuation-dissipation
theorem, we see that
\begin{eqnarray*}
\frac{1}{\widetilde{K}_k} &=& \lim_{v_k\rightarrow 0}
\frac{1}{v_k} \left<
\frac{\partial\phi}{\partial x_k}\right> \\
&=& \frac{1}{V}\int\!\!\int d^2\!{\bf r}d^2{\bf r}'
\left[\left< \frac{\partial\phi({\bf r})}{\partial x_k}
\frac{\partial\phi({\bf r}')}{\partial x_k}\right>_0 -
\left< \frac{\partial\phi({\bf r})}{\partial x_k}\right>_0
\left< \frac{\partial\phi({\bf r}')}{\partial x_k}\right>_0 \right]
\end{eqnarray*}
where $V$ is the volume and $\langle\cdots\rangle_0$ 
denotes the average with $v_k =0$. These quantities are called 
helicity moduli and serve as order parameters for systems exhibiting 
the KT transition, e.g., XY model~\cite{xyhel}.

Using Eqs.~(\ref{ht_n}) and (\ref{cgh}) which
connect the Gaussian model and the domain-wall model, 
$\widetilde{K}_k$ can be written in terms of the quantities of the 
domain-wall model as
$$
\frac{1}{\widetilde{K}_k} = \frac{4\pi^2}{L^2} \sum_{i,j} \left[
\left< \left(\Delta_k n_{i}-\frac{2}{3}\delta_{k,1}\right)
       \left(\Delta_k n_{j}-\frac{2}{3}\delta_{k,1}\right)\right>_0 -
\left< \Delta_k n_{i}-\frac{2}{3}\delta_{k,1}\right>_0
\left< \Delta_k n_{j}-\frac{2}{3}\delta_{k,1}\right>_0
\right]
$$
where $\Delta_k n_{i} \equiv n_{i+{\bf e}_k}-n_{i}$, 
${\bf e}_k$ is the unit vector along the $k$ direction and
$\delta_{k,1}$ is the Kronecker delta function.
From the definition of $n_i$, we see that
\begin{eqnarray*}
\Delta_1 n_{i} &=& \left\{
\begin{array}{ll}
1\ &,\mbox{if there is a domain wall between $(x_i,y_i)$ and
$(x_i,y_i)+{\bf e}_1$} \\
0\ &, \mbox{otherwise}
\end{array} \right. \\
\Delta_2 n_{i} &=& \left\{
\begin{array}{cl}
1\ &,\mbox{if there is a domain wall of type 1 between $(x_i,y_i)$ 
and $(x_i,y_i)+{\bf e}_2$} \\
-1\ &,\mbox{if there is a domain wall of type 2 between $(x_i,y_i)$ 
and $(x_i,y_i)+{\bf e}_2$} \\
0\ &,\mbox{otherwise}
\end{array} \right.
\end{eqnarray*}
Using this relation, we can write the expression for the stiffness
constants as 
\begin{eqnarray}
\frac{1}{\widetilde{K}_1} &=& \frac{4\pi^2}{L^2} \left( \langle
Q^2\rangle-\langle Q\rangle^2\right) \nonumber \\
\frac{1}{\widetilde{K}_2} &=& \frac{4\pi^2}{L^2}
\left< \left(\frac{Q_2-Q_1}{2}\right)^2 \right>
= \frac{4\pi^2}{L^2} \left<\left(\sum_i \Delta X_i\right)^2 \right>
\end{eqnarray}
where $Q\ (Q_{1,2})$ is the total length of the
domain walls (of type 1 or 2) and $\Delta X_i=p_i(L)-p_i(0)$ is the
difference of the position of the $i$-th domain wall at the top to 
that at the bottom. The equations show that $\widetilde{K}_1$ 
controls the fluctuations of the domain-wall length and 
$\widetilde{K}_2$ the fluctuations of the domain-wall 
positions.

We performed simulations with the periodic boundary condition 
along the horizontal direction and the free boundary condition along 
the vertical direction with domain-wall density fixed to $2/3$.
Updating rule from a domain-wall configuration
$\{p_i(j)\}$ to $\{p_i'(j)\}$ is as follows. There are four kinds of 
configurations for any $(i,j)$;
$$
\begin{array}{clllll}
(i)  & p_i(j-1)+1/2 &=& p_i(j) &=& p_i(j+1)-1/2 \\
(ii) & p_i(j-1)-1/2 &=& p_i(j) &=& p_i(j+1)+1/2 \\
(iii)& p_i(j-1)-1/2 &=& p_i(j) &=& p_i(j+1)-1/2 \\
(iv) & p_i(j-1)+1/2 &=& p_i(j) &=& p_i(j+1)+1/2 \ \ \ .
\end{array}
$$
A given configuration can be changed locally only for the cases 
$(iii)$ and $(iv)$. For the case $(iii)$, $p_i(j)$ is updated to 
$p_i'(j)=p_i(j)+1$ with a probability $y^{{\bf L}(\{p'\})-
{\bf L}(\{p\})}$ provided $p_{i+1}(j)\neq p_i(j)-1$, while for the 
case $(iv)$, $p_i(j)$ is updated to  $p_i'(j) = p_i(j)-1$ with a 
probability $y^{{\bf L}(\{p'\})-{\bf L}(\{p\})}$ peovided 
$p_{i-1}(j)\neq p_i(j)-1$. 
For other cases, the configuration is not altered.
Noting that the updating trial at $(i,j)$ has nothing to do with
domain walls $(i\pm 2,j\pm 2)$, we divide the set $\{(i,j)\}$
into four parts depending on whether $i,j$ are even or odd and 
choose randomly one of the four subsets and update 
the whole domain walls in that subset.

The free boundary condition along the vertical direction is necessary 
to calculate the quantities $\sum_i \Delta X_i$. On the other hand,  
one should adopt, in principle, a grand canonical ensemble in which 
the domain-wall number could vary in other to calculate the 
fluctuations of the domain-wall number.  But it poses too much 
technical difficulty, so we measure instead fluctuations of the 
domain-wall number in a fixed rectangular region of size 
$(L/2)\times L$ during simulations of the $L\times L$ system. For
such an ensemble, an elementary probability consideration shows
that $1/\widetilde{K}_1$ should given by
$$ \frac{1}{\widetilde{K}_1} = \frac{16\pi^2}{L^2} \left(
\langle Q'^2\rangle -\langle Q' \rangle^2 \right) $$
where $Q'$ is the number of domain walls inside the $(L/2)\times L$
region.

In Fig.~8, we present results for the energy fluctuations $C$ which 
is defined by
$$
C = \frac{1}{L^2} \left\langle \left[ {\bf L}(\{p\})-\left\langle 
{\bf L}(\{p\}) \right\rangle\right]^2 \right\rangle
$$
for systems of linear size $L=12,36,60,90 \mbox{\ and\ } 
120$. As expected, there is a converging peak instead of a diverging 
peak. The peak position is below the estimated critical interaction 
strength, i.e., at the low-temperature region. This is consistent 
with the behaviors in the XY model where the peak is at the 
high-temperature region.  The KT transition in the XY model 
is the vortex-antivortex unbinding transition and that in the IDW 
model is a spontaneous symmetry-breaking transition. 
The effects of the vortex-antivortex and the symmetry-breaking field 
have a dual relation~\cite{xykt}
$$
{\cal Z}(2\pi K,y_0,y_p) = {\cal Z}(p^2/2\pi K,y_p,y_0)\left(p/2\pi
K\right)^N
$$
where $K$ is the inverse temperature, $y_0$ is the fugacity of the 
vortex and antivortex, and $y_p$ is the strength of the 
symmetry-breaking field.  We see that the high and low temperature 
regions are inverted.

Figs.~9(a) and (b) show the measured values of $1/\widetilde{K}_1$ and
$1/\widetilde{K}_2$, respectively, as a function of $y$ for several
values of $L$.
Though we did not try to extract the quantitative results, they show 
manifest crossover behaviors near $y\sim y_c$. They become smaller 
and smaller in the ordered regions as the system size $L$ increases 
and the points at which the crossover behaviors set in become closer 
and closer to the estimated critical-interaction strength $y_c$. These
behaviors are in accord with Monte Carlo results of the XY
model~\cite{xyhel}. Finally, Fig.~9(c) shows the values of $g$ obtained
from the relation $g=2\pi\sqrt{\widetilde{K}_1\widetilde{K}_2}$. The
coupling constant estimated in this way is in good agreement with that
obtained from the transfer matrix spectra. That the values of $g$ at
$y_c$ fall short of the expected value $9/4$ is attributed to slow
convergence arising from the presence of logarithmic corrections.
From these numerical results, we conclude that the phase transition
belongs to the universality class of the KT transition.

\section{Summary and discussion}\label{sec5}
In summary, we have investigated the phase transitions in the IDW 
model. There are three phases: the $q=0$ and $q=2/3$ C phases and the 
IC phase. The transition between the $q=0$ C phase and the IC phase 
is the PT transition when the interactions between domain walls are 
repulsive or weakly attractive.  
When the interaction is strongly attractive, it becomes the 
first-order phase transition with a discontinuity of the domain-wall 
density. The nature of the transitions are altered because of 
formation of bounded domain-wall states. As to the transition between 
the $q=2/3$ C phase and the IC phase, 
we suggested the free-energy functional is that of the Gaussian 
model with a symmetry-breaking field with the scaling dimension
$X_{3,0}$.
The symmetry-breaking field accounts for effects of the repulsive 
interaction and induces the KT transition from the critical phase 
to the ordered phase. We confirmed 
this scenario by calculating the stiffness constants from the Monte 
Carlo simulations.  From the transfer matrix method, we calculated 
the Gaussian coupling constant in the IC phase and the critical 
interaction strength. Phase boundaries in the $\varepsilon$-$\mu$
plane are also determined from the transfer matrix method.

The domain-wall model is constructed from the TAFIM with the
restriction that three spins on elementary triangles cannot have the 
same sign. If this is allowed the excitation of this type 
constitutes a topological excitation where two-domain walls are 
created or annihilated~\cite{NohK94}. It plays the role of vortex 
and antivortex excitation in the XY model. The scaling dimension of 
the excitation is $X_{0,2}$. Since it is irrelevant when $g>1$, 
it cannot alter the nature of the C-IC transition of the $q=2/3$ C 
phase where $g=9/4$~\cite{Lan83}.

In the PIDW model where the interaction and the chemical potential is
anisotropic the repulsive interaction stabilizes the $q=1/2$ C
phase. But, the phase transition is of the PT type. Thus the 
anisotropy plays an important role in determining the nature of the 
phase transition.
The anisotropy produces an extra term in the free-energy functional 
which couples to the 
differences of the number of domain walls of type 1 and 2.
So the PIDW model would be described by the effective free-energy
functional
$$
{\cal F} = \int d^2\!{\bf r}\ \left[\ \frac{1}{2}K_1\,
\left(\frac{\partial \phi}{\partial x_1}\right)^2 + \frac{1}{2} K_2\,
\left(\frac{\partial \phi}{\partial x_2}-\rho\right)^2+ V_p
\cos\left(p\phi({\bf r})\right)\, \right]
$$
where the symmetry-breaking field accounts for the ordering induced
from domain-wall interactions and the linear term $-\rho
\partial\phi/\partial x_2$ accounts for the effect of the anisotropy.
In this model, the phase transition is induced by the coupling $\rho$.
It induces a domain-wall type excitation and the phase transition is
in the universality class of the PT transition~\cite{Bak82}.

Even though we studied only the equilibrium properties of the IDW 
model in this work, non-equilibrium version is also of interest since 
it describes a driven domain-wall model or the hypercube stacking 
model~\cite{ForT90} and also the charge conduction in the 
charge-density-wave system~\cite{Gru88}. We leave it for further 
study.

\acknowledgments

This work is supported by Korea Science and Engineering Foundation 
through the Center for Theoretical Physics, and also by Ministry of 
Education through Basic Science Research Institute, of Seoul National 
University.

\pagebreak
\appendix
\renewcommand{\theequation}{A\arabic{equation}}
\setcounter{equation}{0}
\section{}
In this appendix, we derive expressions for the largest eigenvalue
$\lambda$ and the corresponding eigenket $|\lambda\rangle$ of the
transfer matrix ${\bf T}_Q$ for $Q=2$ analytically.
The eigenket is of the form
$$
|\lambda\rangle = \sum_{1\leq m <n\leq N} a(m,n) |m,n\rangle
$$
where $|m,n\rangle$ denotes a state with two domain walls at sites 
$m$ and $n$. The eigenvalue equation ${\bf T}_2
|\lambda\rangle=\lambda|\lambda\rangle$ gives linear equations for
coefficients $a(m,n)$:
\begin{equation}\label{eveq}
\lambda a(m,n) = \sum_{m'<n'} a(m',n') 
\langle m,n|{\bf T}_2|m',n'\rangle \ \ \ .
\end{equation}
When ${\bf T}_2$ is applied to a stateket $|m,n\rangle$, it generates
a new stateket:
$$
{\bf T}_2|m,n\rangle = \left\{
\begin{array}{ll}
|m,n\rangle+|m,n+1\rangle+|m+1,n\rangle+|m+1,n+1\rangle& \mbox{if\ }
n>m+1\\
y|m,n\rangle+|m,n+1\rangle+y|m+1,n+1\rangle& \mbox{if\ } n=m+1
\end{array}
\right.
$$
where $y$ is the interaction parameter. Thus Eq.~(\ref{eveq})
becomes
\begin{eqnarray}
\lambda a(m,n\neq m+1) &=& a(m,n)+a(m-1,n-1)+a(m,n-1)+a(m-1,n) 
\label{nneqmp1} \\ 
\lambda a(m,n=m+1) &=& ya(m,m+1)+ya(m-1,m)+a(m-1,m+1) \ \ \ .
\label{neqmp1}
\end{eqnarray}
These equations are solved by the Bethe Ansatz 
\begin{equation}\label{ansatz}
a(m,n) = A_{12} z_1^{-m}z_2^{-n} + A_{21} z_1^{-n}z_2^{-m}
\end{equation}
where $A_{12}$, $A_{21}$, $z_1$ and $z_2$ are constants to be found.
Eq.~(\ref{nneqmp1}) is automatically satisfied by $a(m,n)$ given in 
Eq.~(\ref{ansatz}) provided the eigenvalue is
$$
\lambda = (1+z_1)(1+z_2) \ \ \ .
$$
We can make Eq.~(\ref{neqmp1}) be satisfied by imposing the condition 
that
\begin{eqnarray*}
\left[ a(m,n)+a(m-1,n-1)+a(m,n-1)+a(m-1,n)\right]_{n=m+1} \\ =
ya(m,m+1)+ya(m-1,m)+a(m-1,m+1)\ \ \ .
\end{eqnarray*}
This gives the relation between $A_{12}$ and $A_{21}$ as 
$$
\frac{A_{12}}{A_{21}} = -\frac{
1-(y-1)\left(\frac{1}{z_1}+z_2\right)}{
1-(y-1)\left(z_1+\frac{1}{z_2}\right)} \ \ \ .
$$
If we use a periodic boundary condition $a(n,m+N)=a(m,n)$, we obtain 
the equations for $z_1$ and $z_2$:
$$
z_1^N = \frac{A_{21}}{A_{12}},\ \ \ 
z_2^N = \frac{A_{12}}{A_{21}}\ \ \ .
$$
Multiplying the two equations together gives $(z_1z_2)^N=1$ which
implies that $z_1z_2=\tau$ where $\tau$ is an $N$-th root of unity.
The eigenvector corresponding to
the maximum eigenvalue comes from the solution with $\tau=1$ which
means that $z_1=1/z_2\equiv z$. The value of $z$ is determined from
the equation
\begin{equation}\label{zd}
z^N = - \frac{1-2(y-1)z}{1-\frac{2(y-1)}{z}} \ \ \ .
\end{equation}
The solutions of Eq.~(\ref{zd}) is given by
$$
z=\left\{
\begin{array}{ll}
2(y-1) + O\left(\frac{1}{(2y-2)^N}\right) &\mbox{if\ }
y>\frac{3}{2}\\
1+O\left(\frac{1}{N}\right) &\mbox{if\ } y\leq \frac{3}{2}\ \ \ .
\end{array} \right.
$$
In the limit $N\rightarrow \infty$, this gives the eigenvalues and
eigenvectors presented in Eq.~(\ref{vv2}).

Unfortunately the Bethe Ansatz works only for $Q=2$ and fails when 
$Q\geq 3$. Thus we are not able to get analytic results for general 
$Q$. However, the solution for $Q=2$ provides a useful approximation 
near the $q=0$ C to IC transition as discussed in Sec.~\ref{sec3}.

\pagebreak
\begin{center}
{\Large \bf Table Caption}
\end{center}

\noindent
Table I. {The transfer-matrix calculation of the KT transition point 
of the IDW model. Estimates of the critical interaction strength 
$y_1(N)$ is found by solving the equation $\tilde{g}(N)=9/4$ by the 
Newton iteration method. 
The sequence of $y_1(N)$ is fitted to the form
$y_1(N)=y_2+a/(b+\log{N})$ to obtain a sequence of extrapolated values
$y_2(N)$. $y_c$ is estimated by fitting $y_2(N)$ again to the same
asymptotic form. The error is estimated from the difference between
$y_c$ and $y_2(9)$}

\pagebreak
\begin{center}
{\Large \bf Figure Captions}
\end{center}

\noindent
Fig. 1. {Three ground states of the domain-wall system with repulsive 
interactions when $0<\mu<3\varepsilon$. The broken line  represents a 
triangular lattice and solid line represents sections of the domain 
wall.  Each ground state is obtained by periodic repetitions of each 
unit in the vertical direction.}

\noindent
Fig. 2. {An excited state of the domain-wall system with repulsive
interactions. The small numbers on vertices of a triangular lattice
denote heights $h_i$ of an equivalent TISOS configuration and the 
large numbers in dotted triangles denote the coarse-grained heights
$\tilde{h}_\alpha$.}

\noindent
Fig. 3. {An example of a domain-wall configuration in successive two
rows of a triangular lattice of width $N=6$. The domain-wall
configuration of the lower row is represented by the stateket 
$|1,2,3,5\rangle$ while that of the upper one is represented by 
$|1,3,4,5\rangle$.  The transfer-matrix element $\langle 1,3,4,5|{\bf
T}_{Q=4}|1,2,3,5\rangle$ denotes the Boltzmann weight of the above
configuration and is equal to $y$.}

\noindent
Fig. 4(a). {Estimates of the Gaussian coupling constant $g(N)$ as a
function of $q$ for several values of $y$. 
The symbols $\triangle$, $\Diamond$, $+$,
$\Box$, and $\bullet$ correspond to $N=6$, $9$, $12$, $15$, and $18$,
respectively. The values of $y$ are 0.2, 0.4, 0.6, 0.8, and 1.0
from top to bottom at each values of $q$. The broken line
denotes the line $g=9/4$. (b) Same as in (a) but for the values of 
$y$ 1.0, 1.1, 1.2, 1.3, 1.4, and 1.5 from top to bottom.}

\noindent
Fig. 5. {Phase boundary of the $q=2/3$ C-IC transition in the
$\mu$-$y$ plane. $\mu_{\pm}$ are obtained by extrapolating  
$\mu_{\pm}(N)$. The broken line is drawn at the critical interaction 
strength $y=0.252$. $\mu_+$ and $\mu_-$ merge into a single value 
near that line. Solid lines are guides to the eye.}

\noindent
Fig. 6. {Estimates of $\mu_0$ (a) and $q_0$ (b) obtained from
transfer-matrix calculations (symbols). The two-domain-wall
approximation for $\mu_0$ and the approximation $q_0\sim 1/d_0$ are
also shown in (a) and (b), respectively, by solid lines.}

\noindent
Fig. 7. {Phase diagrams of the IDW model with the repulsive 
$(\varepsilon>0)$ and attractive $(\varepsilon<0)$ interactions. 
The solid lines (---) denote the PT
transition, the broken line (-- -- --) the KT transition and
the dotted line ($\cdots$) the first-order transition. The PT
transition lines are straight lines with slope $-1/\log{2}$.}

\noindent
Fig. 8. {Monte Carlo results for the energy fluctuation $C$. 
This result indicates that the energy 
fluctuation remains finite when the system enters the ordered phase. 
The broken line is drawn at $y=0.252$.}

\noindent
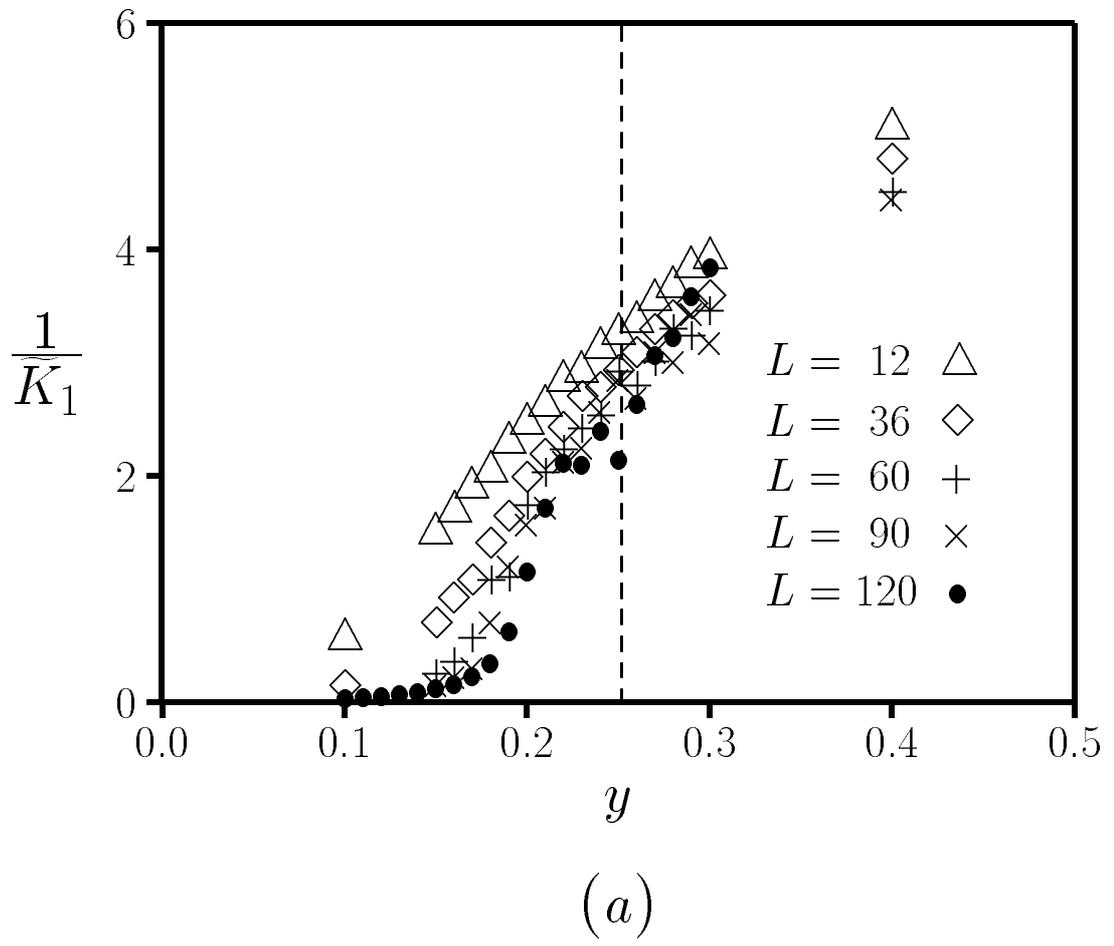
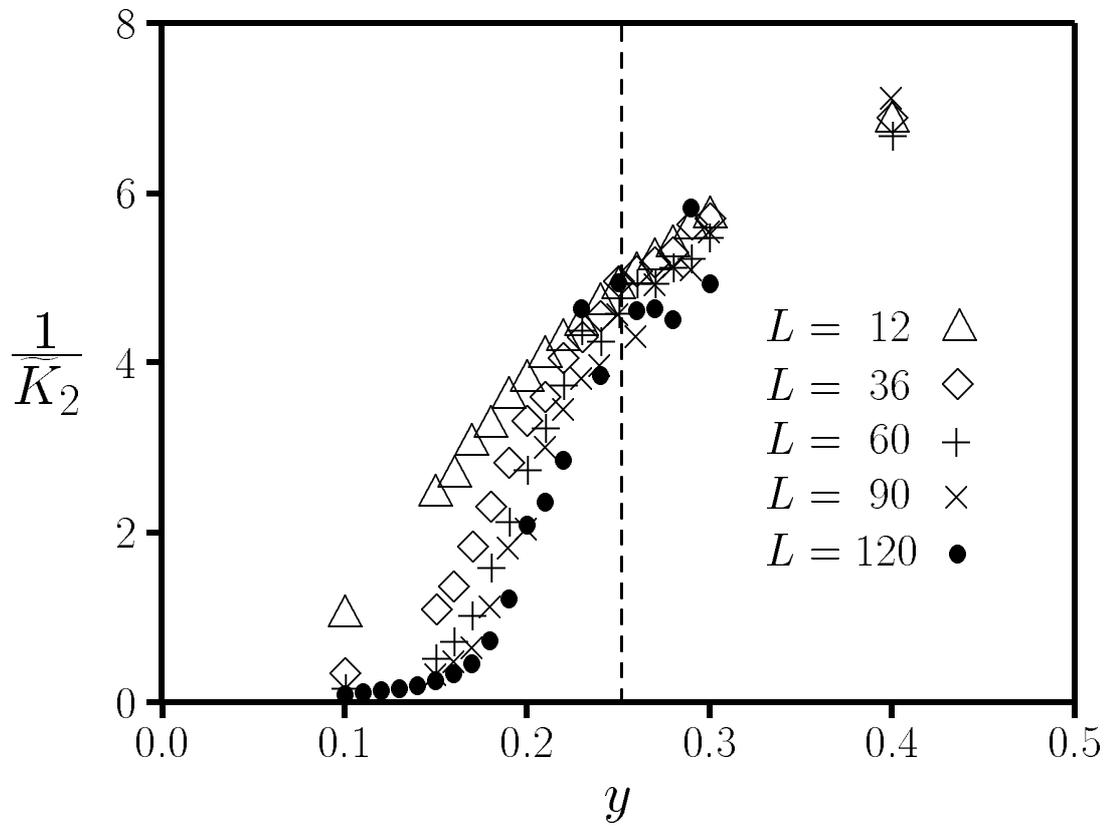
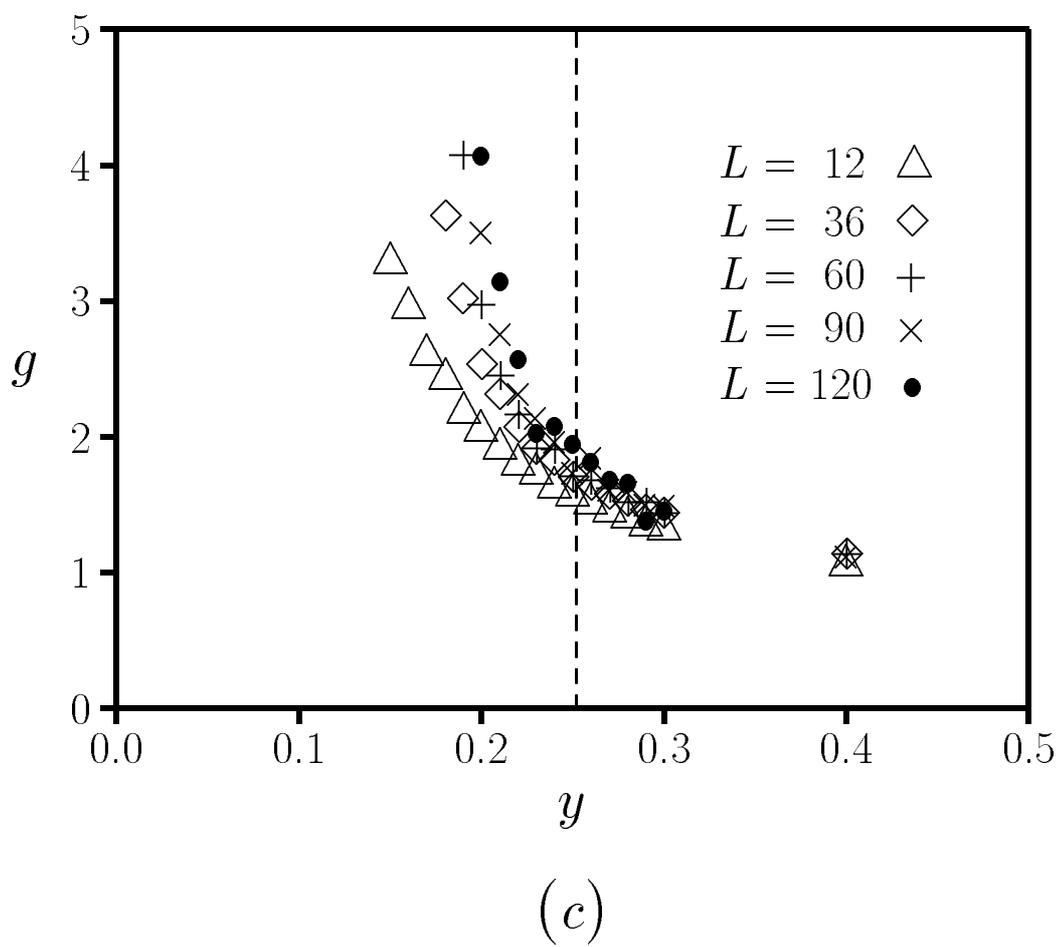
Fig. 9. {Monte Carlo results for the renormalized stiffness constants
$\widetilde{K}_1$ (a),  $\widetilde{K}_2$ (b) and the 
Gaussian coupling constant $g$ (c). They show characteristic behaviors
of the KT transition.}

\begin{table}
\begin{tabular}{cccc} \hline
$N$ & $y_1$          & $y_2$    & $y_c$    \\ \hline
3   & 0.170790486223 & 0.265782 & 0.252    \\
6   & 0.197135051080 & 0.256660 &          \\
9   & 0.206717177287 & 0.255360 &          \\
12  & 0.211836668182 &          &          \\
15  & 0.215121556318 &          &          \\ \hline
\end{tabular}
\end{table}

\vspace{5cm}
\begin{center}
Table~I
\end{center}


\begin{references}
\bibitem{BloH82} H.~W.~J.~Bl\"{o}te and H.~J.~Hilhorst, J.~Phys.~A :
Math. Gen. {\bf 15}, L631 (1982).
\bibitem{NieHB84} B.~Nienhuis, H.~J.~Hilhorst and H.~W.~J.~Bl\"{o}te,
J. Phys. A : Math. Gen. {\bf 17} 3559, (1984).
\bibitem{ForT90} B.~M.~Forrest and L.~-H.~Tang, Phys. Rev. Lett. {\bf
64}, 1405 (1990).
\bibitem{NohK94} J.~D.~Noh and D.~Kim, Phys. Rev. E {\bf 49}, 1943
(1994).
\bibitem{Bak82} P.~Bak, Rep. Prog. Phys. {\bf 45}, 587 (1982).
\bibitem{VilB81} J.~Villian and P.~Bak, J. Phys. (Paris) {\bf 42}, 
657 (1991).
\bibitem{PokT79} V.~L.~Pokrovsky and A.~L.~Talapov, Phys. Rev. Lett.
{\bf 42}, 65 (1979).
\bibitem{BloNWH91} H.~W.~J.~Bl\"{o}te, M.~P.~Nightingale, X.~N.~
Wu and A.~Hoogland, Phys. Rev. B {\bf 43}, 8751 (1991).
\bibitem{BloN93} H.~W.~J.~Bl\"{o}te and M.~P.~Nightingale,
Phys. Rev. B {\bf 47}, 15046 (1993).
\bibitem{xykt} J.~M.~Kosterlitz and D.~J.~Thouless, J. Phys. C {\bf
6}, 1181 (1973); J.~M.~Kosterlitz, J. Phys. C {\bf 7}, 1046 (1974); 
J.~V.~Jose, L.~P.~Kadanoff, S.~Kirkpatrick and D.~R.~Nelson, Phys.
Rev. B {\bf 16}, 1217 (1977).
\bibitem{KimP87} J.~L.~Cardy, {\it Phase transitions and critical
phenomena} vol.~11, edited by C.~Domb and J.~L.~Lebowits 
(Academic Press, 1987);
D.~Kim and P.~A.~Pearce, J. Phys. A {\bf 20}, L451
(1987).
\bibitem{xyhel} C.~Ebner and D.~Stroud, Phys. Rev. B {\bf 28}, 5053
(1983); W.~Y.~Shih, C.~Ebner and D.~Stroud, Phys. Rev. B {\bf 30}, 
134 (1984); H.~Weber and P.~Minnhagen, Phys. Rev. B {\bf 37}, 5986
(1988).
\bibitem{Lan83} D.~P.~Landau, Phys. Rev. B {\bf 27}, 5604 (1983).
\bibitem{Gru88} G. Gr\"{u}ner, Rev. Mod. Phys. {\bf 60}, 1129 (1988).
\end{references}
\end{document}